\documentclass[twocolumn,aps,prb,showpacs,amsmath,superscriptaddress,longbibliography,notitlepage]{revtex4-2}
\usepackage[T1]{fontenc}
\usepackage[latin9]{inputenc}
\usepackage{array}
\usepackage{multirow}
\usepackage{graphicx}
\usepackage{amsmath,amsfonts,amssymb,color}
\usepackage{amsthm}
\usepackage{mathtools}
\usepackage{leftidx}
\usepackage{xcolor}
\usepackage{dcolumn}
\usepackage{bm}
\usepackage{epstopdf}
\usepackage{epsfig}
\usepackage{environ}
\usepackage{pdfcomment}
\usepackage{float}
\usepackage[T1]{fontenc}
\usepackage{setspace}
\usepackage{esint}

\hypersetup{colorlinks=true,citecolor=blue,
linkcolor=blue,urlcolor=blue,pdfstartview=FitH,
bookmarksopen=true}

\begin{document}

\title{Universal fluctuations of localized two interacting particles in one dimension}
\author{Sen Mu}\email{senmu@u.nus.edu}
\affiliation{Centre for Quantum Technologies, National University of Singapore, Singapore 117543, Singapore}
\affiliation{Max Planck Institute for the Physics of Complex Systems, N\"othnitzer Str.~38, 01187 Dresden, Germany.}

\author{Gabriel Lemari\'e}\email{lemarie@irsamc.ups-tlse.fr}
\affiliation{Centre for Quantum Technologies, National University of Singapore, Singapore 117543, Singapore}
\affiliation{Department of Physics, National University of Singapore, Singapore 117542, Singapore}
\affiliation{MajuLab, CNRS-UCA-SU-NUS-NTU International Joint Research Unit, Singapore}
\affiliation{Laboratoire de Physique Th\'{e}orique, Universit\'{e} de Toulouse, CNRS, UPS, France}

\author{Jiangbin Gong}\email{phygj@nus.edu.sg}
\affiliation{Centre for Quantum Technologies, National University of Singapore, Singapore 117543, Singapore}
\affiliation{Department of Physics, National University of Singapore, Singapore 117542, Singapore}
\affiliation{MajuLab, CNRS-UCA-SU-NUS-NTU International Joint Research Unit, Singapore}

\date{\today}

\begin{abstract}
We investigate the universal fluctuations of localized wavefunction in the Fock space of two interacting particles in one-dimensional disordered systems, focusing on the interplay between random potentials and random long-range interactions. By mapping the system onto a directed polymer problem, we show that random potentials alone produce correlated energies for the sites in the Fock space, giving rise to the fluctuation growth exponent $1/2$. Introducing random long-range interactions alters these correlations and drives the system's fluctuations into the Kardar-Parisi-Zhang universality class in (1+1)D with the exponent $1/3$. To validate the universality of the observed fluctuation scaling, we study a complex directed polymer model with competing point and columnar disorder. Our results confirm that columnar disorder corresponds to on-site energies in the Fock space from the random potentials, while point disorder models the effects of random long-range interactions between the two particles. These findings provide new insights into the Fock-space perspective for examining disordered quantum many-body systems,  and emphasize the critical role of disorder structure in determining the universality class of fluctuations in localized quantum systems.
\end{abstract}

\maketitle

\section{Introduction}

Universal fluctuations play a pivotal role in {understanding} diverse aspects of Anderson localization~\cite{Anderson_prl1958,Abrahams_1979,Abrahams_50al}, {manifested} in phenomena such as universal conductance fluctuations~\cite{Lee_1985ucf,Mello_1988prl}, random matrix statistics~\cite{Beenakker_1997rmt}, log-normal distributions in one-dimensional systems~\cite{Mirlin_pr2000}, and multifractal statistics of observables at the critical regime of the Anderson transition~\cite{Lee_rmp1985,Evers_rmp2008,Wegner1980,Castellani_1986,Feigelman2010}. While Anderson localization is fundamentally a single-particle phenomenon, its behavior in the presence of many-body interactions has long been an open and challenging question~\cite{Mott01011967,Anderson80prb}. In particular, the notion of many-body localization (MBL) has sparked renewed and intensive interest~\cite{Huse15annurev,Abanin17rp,Alet18crp,Abanin19rmp,sierant2024}, driven by early studies~\cite{Mirlin05prl,BASKO20061126} showing that interacting fermions in a disordered chain can undergo a metal-insulator transition at weak interactions and finite temperatures. 
{An interesting minimal case is that of two interacting particles (TIP) in a disordered system. The TIP problem has been extensively studied with a delocalization transition observed in two dimensions~\cite{shepelyansky19993tip2d,romer_numerical_1999,ortuno_localized_1999}, while the system remains localized in one dimension~\cite{shepelyansky94,Imry95,weinmann_level_1996,frahm_effective_1996,frahm_two_1997,waintal_two_1998,waintal_two_1999,de_toro_arias_two_1999}.} Numerous studies have focused on how the localization length of the most extended TIP states in 1D scales with the single-particle localization length in the weak disorder regime~\cite{frahm_scaling_1995,borgonovi_enhancement_1995,von_oppen_interaction-induced_1996,Rudolf97prl,ponomarev_coherent_1997,krimer_two_2011,Krimer15prb,frahm_eigenfunction_2016,thongjaomayum_taming_2019}. By contrast, the fluctuation properties of the TIP wavefunction in strongly disordered regimes remain much less explored, representing a significant gap in understanding how interactions influence fluctuations in disordered quantum systems.

{A promising approach to understanding disordered quantum many-body systems is to analyze the many-body wavefunction in the Fock space~\cite{Laflorencie19,Roy20prl,Tarzia2020prb,Laflorencie21,Biroliprb2024,mirlin2024prb,Roy_2024}. This framework allows the dynamical properties of the interacting system to be mapped onto an effective single-particle problem, though the latter is defined on a complex high-dimensional lattice. It is important to emphasize that the disordered quantum many-body problem in the Fock space is fundamentally distinct from the single-particle Anderson localization in higher dimensions~\cite{Tikhonov16,Lemarie17prl}. For instance, by analyzing the one-dimensional disordered TIP problem in {the} Fock space, we can map it onto an equivalent single-particle problem in a two-dimensional disordered lattice. However, different from the genuine two-dimensional Anderson localization, where disordered potentials are typically uncorrelated, the mapped TIP system exhibits strongly correlated potentials~\cite{Welsh_2018,mirlin2024prb}. Introducing random long-range interactions disrupts these correlations, effectively rendering the disorder uncorrelated. 
Thus, the TIP problem with random long-range interactions serves as a minimal model to study the effects of interactions on the fluctuation properties of disordered quantum systems. Notably, the mapped two-dimensional single-particle system offers a natural framework for exploring how correlations in disorder influence fluctuations, an underexplored aspect in the {problem of} two-dimensional Anderson localization.}

The Fock-space view of the TIP problem also provides a resource-efficient alternative to investigate two-dimensional single-particle systems, leveraging the Fock space of the TIP to emulate a two-dimensional spatial lattice. Quantum simulators have emerged as powerful tools for probing quantum phases of matter both in and out of equilibrium, ranging from quantum spin liquids~\cite{Lukin21sci,Roushan21sci} to ergodicity breaking in quantum dynamics~\cite{Bloch15mbl,Bloch16mbl}. These platforms enable in-situ and real-time investigations, making them particularly suited for studying dynamical processes. A wide range of experimental setups~\cite{Blatt2012,Islam13sci,Ritsch13rmp,Lukin16sci,Labuhn2016,Bloch17prx,Mivehvar02012021,Yao21rmp,Bloch22prl}, including Rydberg atoms in optical tweezers, quantum gases in optical cavities, and trapped ions in optical lattices, naturally feature long-range interactions, where the coupling between microscopic constituents decays as a power law with distance {of separation}~\cite{Defenu23rmp}. Many of these platforms also allow for quenched disorder not only in onsite energies but in interaction strengths~\cite{Randall21sci,Mi2022}, greatly expanding their utility for studying disordered quantum systems. Consequently, they provide an ideal avenue for exploring the fluctuation properties of disordered quantum ystems.

In this work, we investigate the universal fluctuation of the localized wave functions in the Fock space of TIP in strongly disordered one-dimensional systems. We focus on how random potentials and interactions jointly govern the universality class of fluctuations. We show that random potential alone induces strongly correlated on-site energies in {the} Fock space, effectively causing the fluctuation scaling to behave as if the system were a direct sum of two independent particles in one-dimensional disordered systems.
However, when random long-range interactions are introduced, these correlations are disrupted, rendering the on-site energies uncorrelated. Consequently, the wave function of the TIP in {its} Fock space encounters a genuinely two-dimensional disordered potential that fundamentally alters its fluctuation scaling. Building on the forward scattering approximation (FSA)~\cite{Stern1973,NSS1985tunnel,Kardar1992,ROS2015,Antonello2016,Baldwin2017} and our previous results on the density fluctuations of localized wave packets in two dimensions~\cite{mu2023kpz}, we demonstrate that the localized wave functions in the Fock space of TIP exhibit the universal fluctuation scaling characteristic of the Kardar-Parisi-Zhang (KPZ) universality class in $(1+1)$ dimensions~\cite{kpz1986,Corwin2012kpz,Halpin2015jsp,Takeuchi2018kpz,Spohn2020kpz}. This underscores the crucial role of random long-range interactions in driving the transition of fluctuation scaling to different universality class.  

The directed polymer (DP) problem in $(1+1)$ dimensions is one of the most known solvable models in the KPZ universality class~\cite{Kardar_prl1987,mezard1990glassy,Zhang1995dp,Calabrese_2010,Dotsenko_2010,PhysRevE.101.012134}. To validate the universality of the observed scaling, we study a variant of the directed polymer model with competing point (uncorrelated) and columnar (correlated) disorder~\cite{Tang1993prl,Balents1993epl,Arsenin1994pre,Kardar1987,Medina1989pra,Chu2016pre}, where a similar transition is observed. This confirms that disorder correlations play an essential role in determining the fluctuation scaling. Our findings establish TIP with random long-range interactions as an ideal minimal model for investigating how disorder correlations influence fluctuation scaling in quantum disordered systems. Furthermore, this work provides a pathway to study fluctuations in quantum many-body localized systems, highlighting the impact of random long-range interactions through the fluctuation scaling of the wavefunction in the Fock space.

Our paper is organized as follows: In Section~\ref{model}, we introduce our model and its corresponding Fock-space graph. In Section~\ref{results}, we apply the forward scattering approximation to predict the fluctuation scaling behavior of the localized wavefunction in the Fock space, examining cases with and without random long-range interactions for the TIP. In Section~\ref{DP}, we compare our findings with a variant of the directed polymer model with competing columnar and point disorders, thereby confirming the universality of our findings. Finally, in Section~\ref{conclusion}, we conclude our study and outline promising directions for future exploration.

\section{TIP MODEL AND ITS FOCK SPACE}
\label{model}

We start by introducing the model for the disordered TIP problem, focusing on the key features of their corresponding Fock-space graphs.

\subsection{Disordered two interacting spinless fermions}
{Let us consider interacting spinless fermions in a disordered chain under open boundary condition (OBC)}, described by the {following} Hamiltonian:

\begin{eqnarray}
H = &-& J \sum_{i=1}^{L-1} \left( c_i^\dagger c_{i+1} + {\rm h.c.} \right)\nonumber\\ 
&+& \sum_{i=1}^{L} V_i \hat{n}_i + \sum_{i<j}^{L} \frac{\Delta_{i,j}}{|i-j|^\alpha} \hat{n}_i \hat{n}_j
\label{eq:dis_tip_model}
\end{eqnarray}
where $c^\dagger_i$ ($c_i$) are the fermionic creation (annihilation) operators, and $\hat{n}_i = c^\dagger_i c_i$ is the number operator at site $i$. We set the hopping strength $J=1$. The random potential $V_i$ and the density-density interaction strength $\Delta_{i,j}$ are independently and identically drawn from uniform distributions $[-W_V/2, W_V/2]$ and $[-W_D/2, W_D/2]$, respectively, with $W_V$ and $W_D$ representing the corresponding disorder strengths. The decay of the pairwise interaction terms is controlled by the power-law exponent $\alpha$. {For $\alpha>1$, the system is weakly long-range interacting and approaches short-range behavior as $\alpha$ increases. In contrast, for $\alpha < 1$, the system is strongly long-range interacting~\cite{Defenu23rmp}. Especially, when $\alpha = 0$, all pairs of fermions interact equally.}

With the Jordan-Wigner transformation, {such a} spinless fermion model can be mapped onto {a} disordered XXZ spin chain up to an additive constant,
\begin{eqnarray}
H_{\rm XXZ} = &-&\frac{J}{2}\sum_i^{L-1}({\sigma}_i^x{\sigma}^x_{i+1}+{\sigma}^y_i{\sigma}^y_{i+1})\nonumber\\
&+& \sum_i^L\frac{V_i}{2}\sigma^z_i+\sum_{i<j}^{L}\frac{\Delta_{i,j}}{4|i-j|^\alpha}{\sigma}^z_i{\sigma}^z_j,
\label{eq:xxz}
\end{eqnarray}
where $\vec{\sigma_i}=(\sigma_i^x,\sigma_i^y,\sigma_i^z)$ are the Pauli matrices acting on {a} spin-1/2 degree of freedom {associated with} each lattice site. 

The Hamiltonian in Eq.~(\ref{eq:dis_tip_model}) conserves the total number of spinless fermions, which corresponds to the conservation of total $z$-magnetization in the XXZ model described by Eq.~(\ref{eq:xxz}). For the TIP problem, we focus on the case where $N = \sum_i n_i = 2$, or equivalently $S^z_{\rm tot} = \sum_i \sigma_i^z = 4 - L$ in the spin representation. Our results shown for the spinless fermions are directly translatable to the spin-1/2 chain in  Eq.~(\ref{eq:xxz}), which can be more experimentally accessible in certain quantum simulators, such as Rydberg atom platforms.

\begin{figure}
\includegraphics[width=1.0\linewidth]{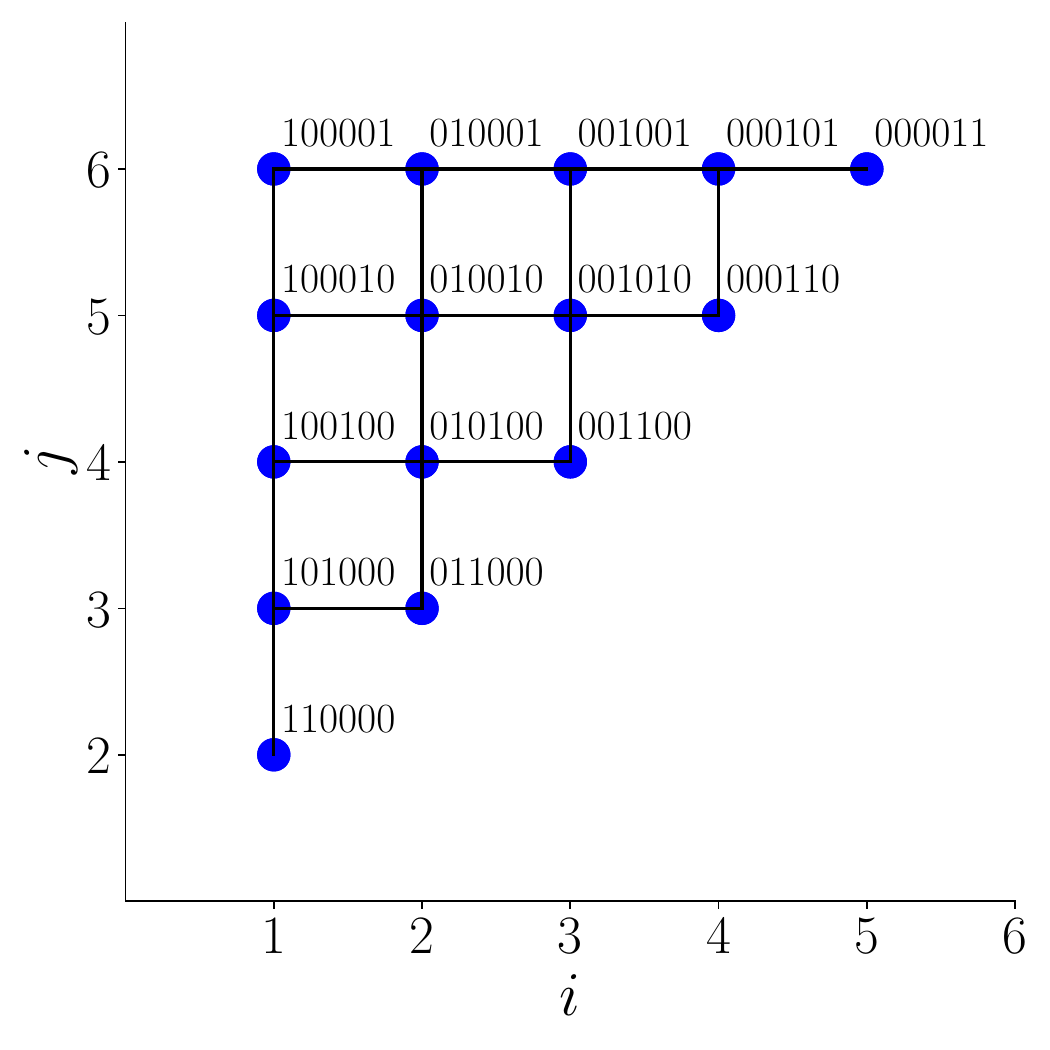}
\caption{Illustration of the Fock space for two spinless fermions on a one-dimensional lattice of size $L=6$. Blue dots represent the basis states of the Fock space, corresponding to different occupation configurations, where $i$ denotes the position of one fermion and $j$ denotes the position of the other. Black lines indicate the one-particle hopping terms that connect different basis states.}
\label{fig:fock_space}
\end{figure}

\subsection{The Fock space of TIP}

Let us now {focus our attention} on the TIP Hamiltonian in the Fock space representation~\cite{Laflorencie19,Roy20prl,Tarzia2020prb,Laflorencie21,Roy_2024}. The Fock space graph consists of sites representing valid two-particle configurations and edges corresponding to hopping terms between these configurations. The number of sites defines the Fock-space dimension $N_F$, and for TIP, $N_F = \binom{L}{2}$. An effective tight-binding Hamiltonian for a fictitious single particle can be written for the TIP system in the Fock space as:

\begin{eqnarray}
H_{\rm Fock} &=& \sum_{S} E_S |S\rangle\langle S| + \sum_{S \neq K} T|S\rangle\langle K|\nonumber\\
&=& H_S + H_T
\label{eq:fock_space}
\end{eqnarray}
where $|S\rangle$ are the sites of the graph shown in Fig.~\ref{fig:fock_space}, corresponding to two-particle states in the configuration basis, e.g., $|100001\rangle$, with associated on-site energies $E_S$. {The} {non-zero} {off-diagonal matrix elements $T$ in $H_{\rm Fock}$} arise from the hopping terms in the TIP Hamiltonian, generating the edges of the graph and acting as hopping amplitudes for the fictitious single particle. {Despite the long-range interactions, only nearest-neighbor hopping occurs in the Fock space. This means that the dynamics viewed from the Fock space are governed by local transitions between configurations connected by single-particle hopping processes, providing a detailed microscopic description of the system.} 

Consequently, we can express the wavefunction for the fictitious particle in the Fock space as

\begin{equation}
|\psi\rangle = \sum_{S=1}^{N_F} \psi_S |S\rangle,
\label{eq:quantum_state}
\end{equation}
where $\psi_S$ represents the amplitude of the wavefunction at the site $|S\rangle$. Due to the indistinguishability of the spinless fermions in our model, Eq.~\eqref{eq:fock_space} corresponds to a tight-binding Hamiltonian defined on the upper triangular portion of a two-dimensional square lattice, with diagonal entries excluded where $i = j$ due to the Pauli exclusion principle. More importantly, the on-site energies 

\begin{equation}
E_S = \sum_{i} V_i \langle S|\hat{n}_i|S\rangle + \sum_{i<j}\frac{\Delta_{i,j}}{|i-j|^\alpha} \langle S|\hat{n}_i \hat{n}_j|S\rangle
\label{eq:energies}
\end{equation}
are inherently correlated in the absence of random long-range interactions ($W_D=0$). This is evident from the fact that the number of distinct $E_S$ values is $L(L-1)/2$, while there are only $L$ independent variables, namely the $V_i$. This underscores the nontrivial role of random long-range interactions, $\Delta_{i,j}$, which can disrupt these correlations, render the $E_S$ uncorrelated, and thereby alter the resulting fluctuation behavior.

\begin{figure}
\includegraphics[width=1.0\linewidth]{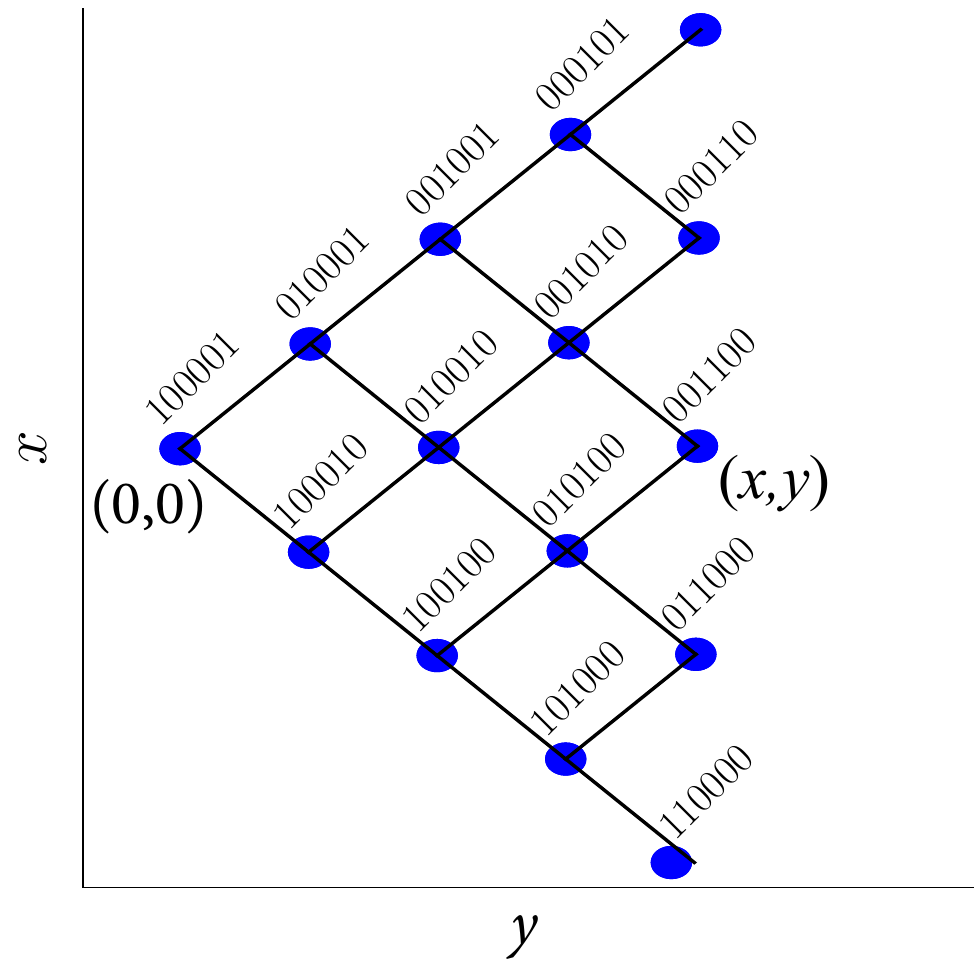}
\caption{A directed polymer (DP) problem mapped from the Fock space of TIP in a one-dimensional lattice of size $L=6$ under the forward scattering approximation. The TIP is initialized at $|Q\rangle$, corresponding to the site $(x=0,y=0)$ in the DP problem, representing the point boundary condition of the DP problem. The distance $r$ in the Fock space is mapped to the time $y$ in the DP problem, while the wavefunction density $|\psi_P(y)|^2$ corresponds to the squared polymer partition function $|Z(x,y)|^2$, with $|P\rangle$ denoting the site $(x=0,y=4)$ in the DP problem.}
\label{fig:fock_space_to_dp}
\end{figure}

\section{Forward Scattering Approximation in the Fock space and proposed form for the fluctuation scaling}
\label{results}

In this section, we first discuss the forward scattering approximation (FSA)~\cite{Stern1973,NSS1985tunnel,Kardar1992,ROS2015,Antonello2016,Baldwin2017} in the Fock space, and then present results on the fluctuation scaling of the localized wavefunction in the Fock space with and without random long-range interactions. Specifically, we initialize the system in a state localized on a single site in the Fock space, corresponding to the site at the top-left corner in Fig.~\ref{fig:fock_space},

\begin{equation}
|\psi^{t=0}\rangle = |Q\rangle = |10\cdots 01\rangle.
\label{eq:initial_state}
\end{equation}
This state corresponds to the configuration where the two particles are maximally separated, located at the ends of the chain. We will explain the reason for this choice of the initial state later. The system is then evolved with the Hamiltonian $H$ given in Eq.~\eqref{eq:dis_tip_model}, both with and without random long-range interactions:

\begin{equation}
|\psi^{t}\rangle = e^{-iHt}|Q\rangle,
\label{eq:evolved_state}
\end{equation}
where we have set $\hbar = 1$. Before presenting our numerical results, we first explore the FSA, a valuable approximation in the strong disorder regime, to gain some insights into the universal fluctuations in the localized wavefunction.

\subsection{Forward scattering approximation in the Fock space}

Let us consider the evolution of the initial state over a long time in a strongly disordered system, until the envelope of the wavefunction exhibits exponential decay from the initial site to distant sites. The notion of distance $r$ in the Fock space is defined as the graph distance in our TIP problem, as illustrated in Fig.~\ref{fig:fock_space}, i.e., the minimum number of edges connecting two sites. 

{The non-local propagator between the site $P$ at a distance $r > 0$ from the initial site $Q$ in the Fock space is given by the matrix element of the Green's function:
\begin{equation}
G_{P,Q}(\omega) = \langle P| \frac{1}{\omega - H + i 0^+} |Q\rangle.
\label{eq:green}
\end{equation}
Using the spectral decomposition, it can be expressed as 
\begin{equation}
G_{P,Q}(\omega) = \sum_X \frac{\psi_X(P)\psi^*_X(Q)}{\omega - E_X + i 0^+},
\label{eq:spectral}
\end{equation}
where $\psi_X$ is the eigenstate with eigenenergy $E_X$, and the amplitude $\psi_X(S)$ is given by $\psi_X(S) = \langle S|\psi_X\rangle$. Thus, the residue at energy $E_X$ is:
\begin{equation}
\lim_{\omega \to E_X} (\omega - E_X) G_{P,Q}(\omega) = \psi_X(P)\psi^*_X(Q).
\label{eq:residue}
\end{equation}
In the strong disorder regime, where the on-site energies $E_S$ in the Fock space are large random values and the hopping amplitude $T$ is small (i.e., $T/W \ll 1$, with $W = \max(W_V, W_D)$ in Eq.~(\ref{eq:fock_space})), the forward scattering approximation (FSA) can be applied to approximate the eigenfunctions~\cite{Stern1973,NSS1985tunnel,Kardar1992,ROS2015,Antonello2016,Baldwin2017}. The FSA treats the hopping term perturbatively to leading order in powers of $T/W$, retaining only the shortest (directed) paths between two sites on the Fock-space graph while discarding longer paths, as the contribution of a path decreases exponentially with its length.} 

{The resulting amplitude of the wavefunction at site $P$ to the lowest order in $E_X\rightarrow E_Q$ and $\psi_X(Q)\rightarrow 1$ under the FSA is given by~\cite{Antonello2016}:
\begin{equation}
\psi_X(P) = \sum_{D\mathcal{P}:Q\rightarrow P} \prod_{l=1}^{r} \frac{T}{E_Q - E_{S_l}},
\label{eq:wave_amp}
\end{equation}
where $D\mathcal{P}$ represents the set of directed paths of length $r$ in the Fock space that connect sites $P$ and $Q$, and $E_{S_l}$ denotes the on-site energy encountered at a site along one of these directed paths. The FSA provides an insightful and efficient approximation for the eigenstates in the strong disorder regime, and eigenstates are expected to reflect the infinite time behavior of quench dynamics. In our system, where the initial state $|Q\rangle$ predominantly overlaps with the eigenstate localized on the same site, i.e., $\psi_X(Q) \approx 1$, the FSA motivates us to focus exclusively on the contributions from the dominant directed paths. This approximation offers valuable insights into the universal fluctuations of the two-dimensional localized wavefunctions observed after long-time evolution~\cite{mu2023kpz}.}

Remarkably, rewriting $T/(E_Q-E_{S_l}) = Te^{-\tilde{V}_l}$ with $\tilde{V}_l=\ln (E_Q-E_{S_l})$, the expression in Eq.~\eqref{eq:wave_amp} closely resembles the partition function of a directed polymer (DP) with complex on-site disorder~\cite{Zhang_prl_complex_dp,Zhang_epl_complex_dp,Medina1989,Kardar1992,Antonello2016}:

\begin{equation}
Z(x,y) = \sum_{\mathcal{DP}:(0,0)\rightarrow (x,y)} \prod_{l=1}^{r} e^{-V(x_l,y_l)} \;,
\label{eq:z_dp}
\end{equation}
where $Z(x,y)$ represents a sum over all possible directed paths $\mathcal{DP}$ connecting $(0,0)$ to $(x,y)$, $V(x_l, y_l)$ denotes the on-site disorder at site $(x_l, y_l)$ along a directed path, and the temperature is incorporated into the rescaling of the on-site disorder. In (1+1) dimensions, numerical evidence shows that the free energy scaling of complex DP problem with complex Boltzmann weights are identical to those of real DP with real weights~\cite{Medina1989,GELFAND199167,BLUM1992588,Roux_1994,PhysRevLett.111.026801}, confirming their belonging to the KPZ universality class.
In Fig.~\ref{fig:fock_space_to_dp}, we illustrate the mapped directed polymer problem from the Fock space of TIP under the FSA. This motivates our choice of the initial state to match the point-to-point boundary condition of the DP problem.

\begin{figure*}
\includegraphics[width=1.0\linewidth]{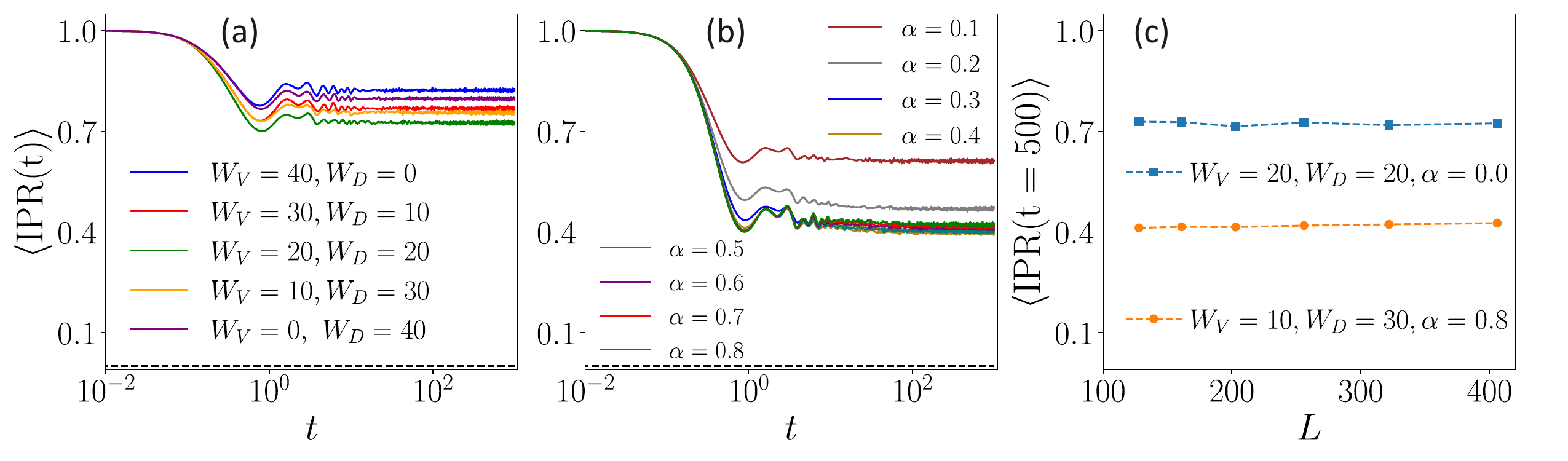}
\caption{Disorder-averaged IPR as a function of time. (a) for cases with varying strengths of random potential and random long-range interactions with $\alpha=0$ (all particle pairs interacting with nonzero $W_D$) and (b) for cases with varying exponent $\alpha$ for the decay of the random long-range interactions with $W_V=10$, $W_D=30$. For $t>100$, the IPR stabilizes at a steady value, indicating that the wavefunction remains localized for the disorder strength considered. The black dashed line represents the IPR for extended states, $1/N_F$, in the Fock space with dimension $N_F = L(L-1)/2$. {(c) The IPR values at $t = 500$ are independent of the system size $L$, confirming the localized phase.} Numerical simulations are performed on a chain of size $L=256$ for (a) and (b) with two spinless fermions, and results are averaged over 1920 disorder realizations.}
\label{fig:IPR_vs_time}
\end{figure*}

\subsection{Proposed form for the fluctuation scaling}

The density of the wave function $|\psi^t(r)|^2$ numerically obtained from Eq.~\eqref{eq:evolved_state} exhibits time dependence, characterized by {persistent temporal fluctuations} in the disordered quantum system~\cite{Ziraldo12prl,Ziraldo13prb}. While the implications of these fluctuations will be addressed in a separate work, we focus here on the sample-to-sample fluctuations of $|\psi^t(r)|^2$ at a fixed large time. In the following, we set the evolution time $t=500$ to analyze the fluctuation scaling of the localized wavefunction in the Fock space. To justify this choice, we present the dynamics of the inverse participation ratio (IPR), defined in the Fock space as:

\begin{equation}
{\rm IPR}(t) = \sum_S |\psi_S(t)|^4.
\end{equation}
IPR quantifies the degree of localization of the wavefunction, with larger values indicating stronger localization. For our initial state at a single site, ${\rm IPR} = 1$, while for an extended state, ${\rm IPR} = 1/N_F$ with $N_F = L(L-1)/2$ the dimension of the Fock space for the TIP problem.

{As shown in Fig.~\ref{fig:IPR_vs_time}(a) and (b), for various parameters relevant to our subsequent investigations, the IPR stabilizes at a (quasi)-steady value for $t>100$, indicating that the wavefunction ceases to propagate further for the disorder strengths considered. Additionally, we verified that the IPR values at $t = 500$ for two parameter sets are independent of the system's size, as presented in Fig.~\ref{fig:IPR_vs_time}(c). This stabilization of the IPR justifies our choice of $t = 500$ as the evolution time for studying fluctuations of the localized wavefunctions, ensuring that the results reflect the properties in the localized phase rather than transient dynamics. For simplicity, we omit the time label from now on for the time-evolved localized wavefunction, i.e., denoting $|\psi_{\rm loc}(r)|^2 \coloneq |\psi^{t=500}(r)|^2$.}

Building on the insights from {the} previous section, we investigate the fluctuation scaling of the localized wavefunction of the TIP in the Fock space as a function of the distance $r$, both with and without random long-range interactions. Comparing with the DP problem, we interpret the logarithmic density of the localized wavefunction, $\ln|\psi_{\rm loc}(r)|^2$, as the free energy of the DP problem. Based on this analogy, we anticipate the following behavior~\cite{mu2023kpz,PhysRevB.43.10728,Takeuchi2011,monthus2012random,Somoza_prl2007,Gabriel_prl2019}:

\begin{equation}
\ln|\psi_{\rm loc}(r)|^2 \displaystyle \underset{r\gg \xi}{\approx} -\frac{2r}{\xi} + \left(\frac{r}{\xi}\right)^\beta \Gamma \, \chi(\boldsymbol{r}) + \Lambda,
\label{Eq:stat_dis}
\end{equation}
where $\chi(\boldsymbol{r})$ is a random variable of order one, and $\Gamma$ and $\Lambda$ are constants. The first term represents exponential localization with $\xi$ denoting the localization length {in the Fock space}, while the second term captures sample-to-sample fluctuations, characterized by the fluctuation growth exponent $\beta$. It is found numerically $\beta \approx 1/3$ for the localized wave packets in two-dimensional Anderson localization, supporting its belonging to the KPZ universality class~\cite{mu2023kpz}.

\section{Numerical results on fluctuation scaling with competing random potentials and random long-range interactions}

\begin{figure}
\includegraphics[width=1.0\linewidth]{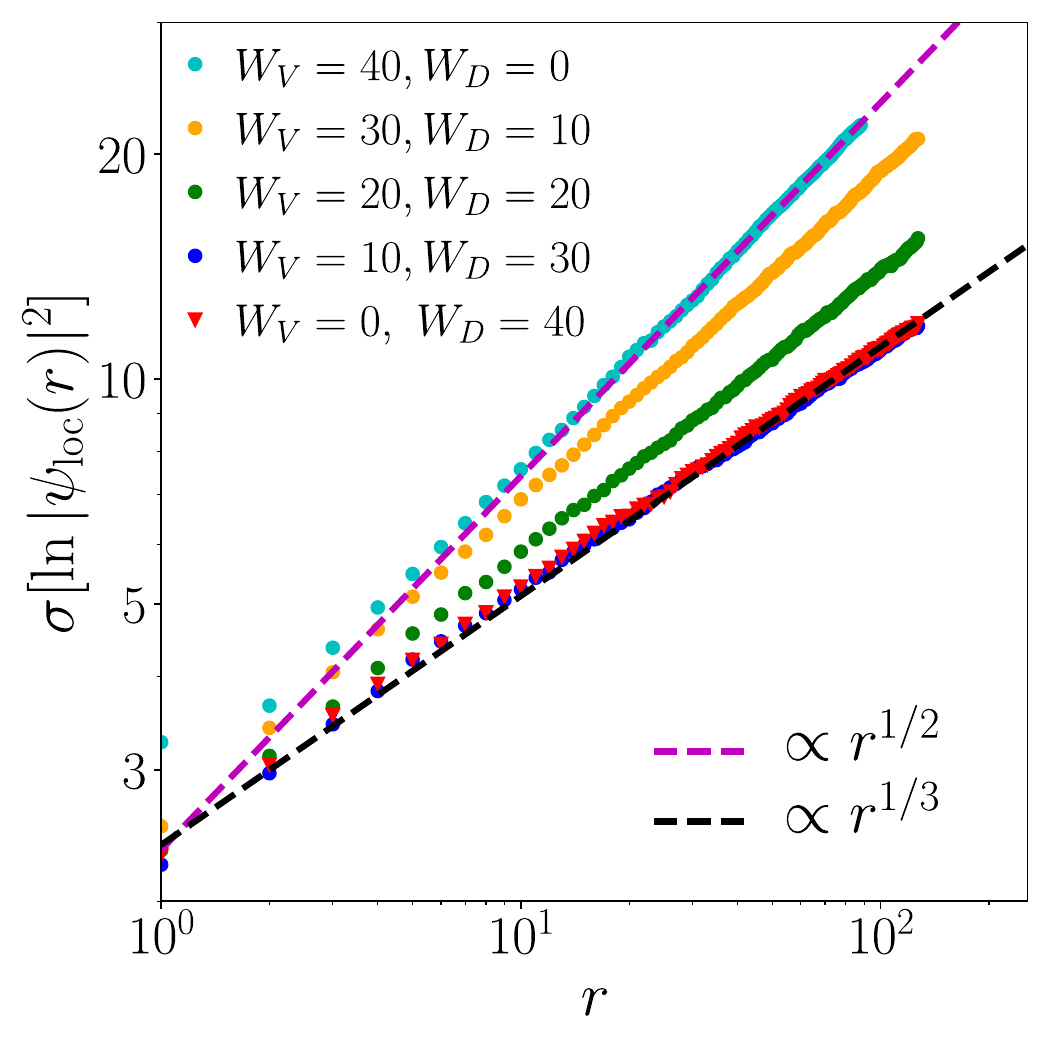}
\caption{Scaling of fluctuations $\sigma[\ln|\psi_{\rm loc}(r)|^2]$ with the graph distance $r$ in the Fock space for different disorder strengths $W_V$ and $W_D$. At each $r$, the selected state is the one horizontally aligned with the initial site, as illustrated in Fig.~\ref{fig:fock_space_to_dp}. The purple and black dashed lines indicate the algebraic behaviors $r^{1/2}$ and $r^{1/3}$, respectively. Note that the data for $W_V=10$ and $W_D=30$ are overlapped by that for $W_V=0$ and $W_D=40$. Numerical simulations are performed with two spinless fermions on a chain of size $L=256$, the evolution time $t=500$, and averaged over $9620$ disorder realizations.}
\label{fig:std_log_P_vs_r}
\end{figure}

\begin{figure*}
\includegraphics[width=1.0\linewidth]{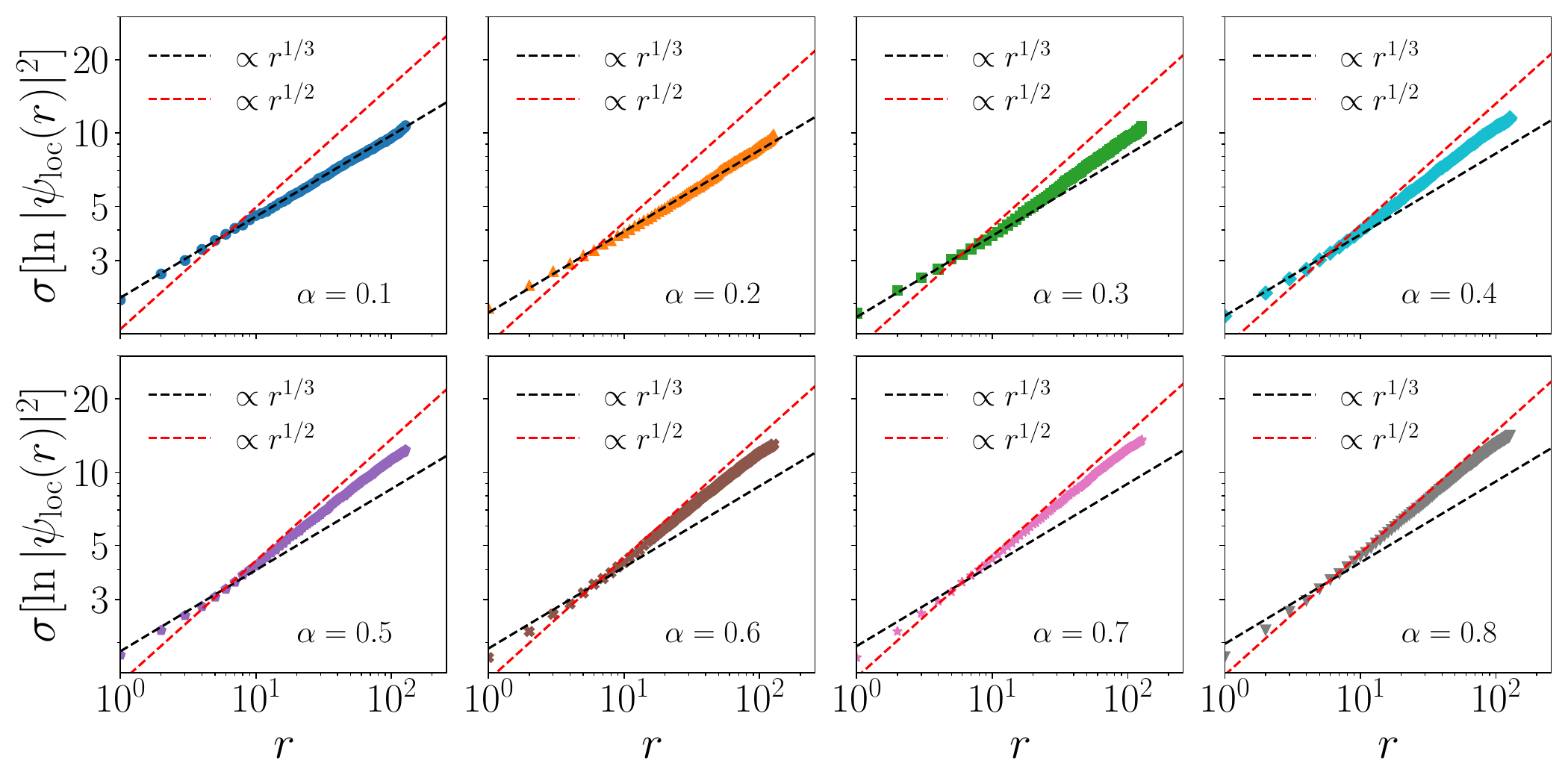}
\caption{Scaling of fluctuations $\sigma[\ln|\psi_{\rm loc}(r)|^2]$ with the graph distance $r$ in the Fock space for different interaction decay exponent $\alpha$. At each $r$, the selected state is the one horizontally aligned with the initial site, as illustrated in Fig.~\ref{fig:fock_space_to_dp}. The red and black dashed lines indicate the algebraic behaviors $r^{1/2}$ and $r^{1/3}$, respectively. Numerical simulations are performed with two spinless fermions on a chain of size $L=256$, disorder strengths $W_V=10$ and $W_D=30$, the evolution time $t=500$, and averaged over $9620$ disorder realizations.
}
\label{fig:std_psi_alpha_Fock}
\end{figure*}

There are two equivalent approaches to implement the numerical simulations to obtain $\ln|\psi_{\rm loc}(r)|^2$. In the first approach, we consider a system with fixed size $L$ and collect the probability of observing a series of target states, such as $|010010\rangle$ and $|001100\rangle$, as shown in Fig.~\ref{fig:fock_space_to_dp}. These states are horizontally aligned to the initial state with varying distances $r$, corresponding to the point-to-point boundary condition in the DP problem. In the second approach, we vary the system size while always initializing the system in the state $|10\ldots01\rangle$ and collect the probability of observing the bound state at the center of the system, $|0\ldots0110\ldots0\rangle$, where the distance $r$ is linearly proportional to the system size $L$. 

For efficiency in numerical simulations, we employ a system with a fixed size $L$ and investigate the fluctuations of observing the target states through real-space observables. Note that the probability of finding certain fermion/spin configurations is experimentally feasible with single-site readout capabilities available on various modern quantum simulators~\cite{Yao21rmp,Randall21sci,Mi2022}.

\subsection{All pairs of particles interacting equally, \texorpdfstring{$\alpha=0$}{alpha=0}}

{Let us now examine how the interplay between random potentials and random long-range interactions influences the fluctuation scaling of the localized wavefunction in the Fock space. Note that the pairwise interaction terms are controlled by the power-law exponent $\alpha$, and we start with the case $\alpha=0$, where particle pairs interact equally regardless of their distances.} Two limiting cases are noteworthy: (i) when $W_D = 0$, the disorder structure in the Fock space is induced solely by the random potential, resulting in correlated on-site energies (columnar disorder); and (ii) when $W_V = 0$, the random long-range interaction manifests as uncorrelated on-site energies (point disorder). To explore the competition between these two types of disorder on the Fock space, we systematically vary their strengths. Starting with $W_V = 40$ and $W_D = 0$, we gradually decrease $W_V$ while increasing $W_D$, eventually reaching $W_V = 0$ and $W_D = 40$.

Fig.~\ref{fig:std_log_P_vs_r} illustrates the behavior of the standard deviation for $\ln|\psi_{\rm loc}(r)|^2$ growing algebraically with the graph distance $r$ as 
\begin{equation}
\sigma[\ln|\psi_{\rm loc}(r)|^2] \sim r^{\beta} \, .
\end{equation}
Here, the standard deviation is defined as $\sigma[X] = \sqrt{\langle X^2 \rangle - \langle X \rangle^2}$ with $\langle X \rangle$ denoting the disorder average of $X$. We also note that for each $r$, we have selected the state that is horizontally aligned with the initial site $Q$ or $(0,0)$, as illustrated in Fig.~\ref{fig:fock_space_to_dp}. In the absence of random long-range interactions, i.e. $W_D=0$, the on-site energy $E_S$ on the Fock space has strong correlations built in, causing the fluctuation exponent to be $\beta \approx 1/2$. This reflects the behavior of an effectively one-dimensional Anderson localization~\cite{Mirlin_pr2000} rather than a generic two-dimensional Anderson localization~\cite{mu2023kpz}.

Remarkably, as the strength of the random long-range interaction $W_D$ increases, the fluctuation scaling exhibits a crossover from $\beta \approx1/2$ to $\beta \approx 1/3$, as shown in Fig.~\ref{fig:std_log_P_vs_r}. This transition highlights the interplay between the correlated columnar disorder from the random potential and the uncorrelated point disorder from the random long-range interactions. As the influence of the long-range interaction grows, the fluctuation scaling undergoes a crossover from the behavior of an effectively one-dimensional Anderson localization, characterized by a fluctuation growth exponent $1/2$, to a generic two-dimensional Anderson localization belonging to the KPZ universality class with $1/3$. Specifically, for $W_V = 10$ and $W_D = 30$, or for $W_V = 0$ and $W_D = 40$, it is clear that $\beta \approx 1/3$, indicating that the uncorrelated point disorder from the long-range interactions dominates over the correlated columnar disorder from the random potential. This crossover underscores the critical role of the relative strengths of different disorder types in shaping the universality class of fluctuation scaling, offering valuable insight into the disorder structures in the Fock space of quantum interacting systems.

\subsection{Interaction strength decaying as a power law with the distance, \texorpdfstring{$\alpha>0$}{alpha>0}}

In previous discussions, we considered the case where all pairs of fermions interact equally but did not address interactions with strength decaying as a power law with a finite $\alpha > 0$. Here, we investigate whether the same fluctuation scaling can be observed with power-law decaying random interactions. Before delving into the effects of $\alpha$ on the fluctuation scaling, we qualitatively examine how the on-site energy $E_S$ in Eq.~(\ref{eq:energies}) is influenced by $\alpha$ in the Fock space. The contribution from the interaction to $E_S$ is always uncorrelated because each particle pair contributes an independent random variable, and the total number of such pairs equals the number of sites in the Fock space for TIP. However, when $\alpha$ is finite, the interaction strength depends on the distance between the two particles: the farther apart the particles are, the less contribution from the interaction to the energy of the corresponding site in {the} Fock space. As $\alpha$ increases from 0, the initial state, where the two particles are far apart, experiences the weakest disordered interaction strength. As the state evolves and the particles move closer together, {more contributions from the interaction to the site energy should be expected}. 

Presented in Fig.~\ref{fig:std_psi_alpha_Fock}, we observe a crossover in the fluctuation scaling as $\alpha$ increases for the system size and disorder strength considered. Specifically, the scaling transitions from $\beta \approx 1/3$ to $\beta \approx1/2$ when $\alpha$ increases from 0 to 0.8. Remarkably, the KPZ exponent $1/3$ persists up to $\alpha = 0.2$ in our numerical simulations. This behavior can be understood by considering the relative strength of the interaction to the random potential. If we set $W_D / L^\alpha \geq W_V$, such that the weakest random interaction contributes a comparable level of disorder as the random potential, we expect point disorder to dominate in the Fock space. In our case, this condition implies $\alpha \leq (\ln W_D - \ln W_V)/\ln L\approx 0.2$, aligning well with our numerical results for the parameters considered $W_V=10, W_D=30$ and $L=256$. We observed that this criterion holds consistently for several other values of $L$.

\section{Analogy to complex directed polymer}
\label{DP}

\begin{figure}
\includegraphics[width=1.0\linewidth]{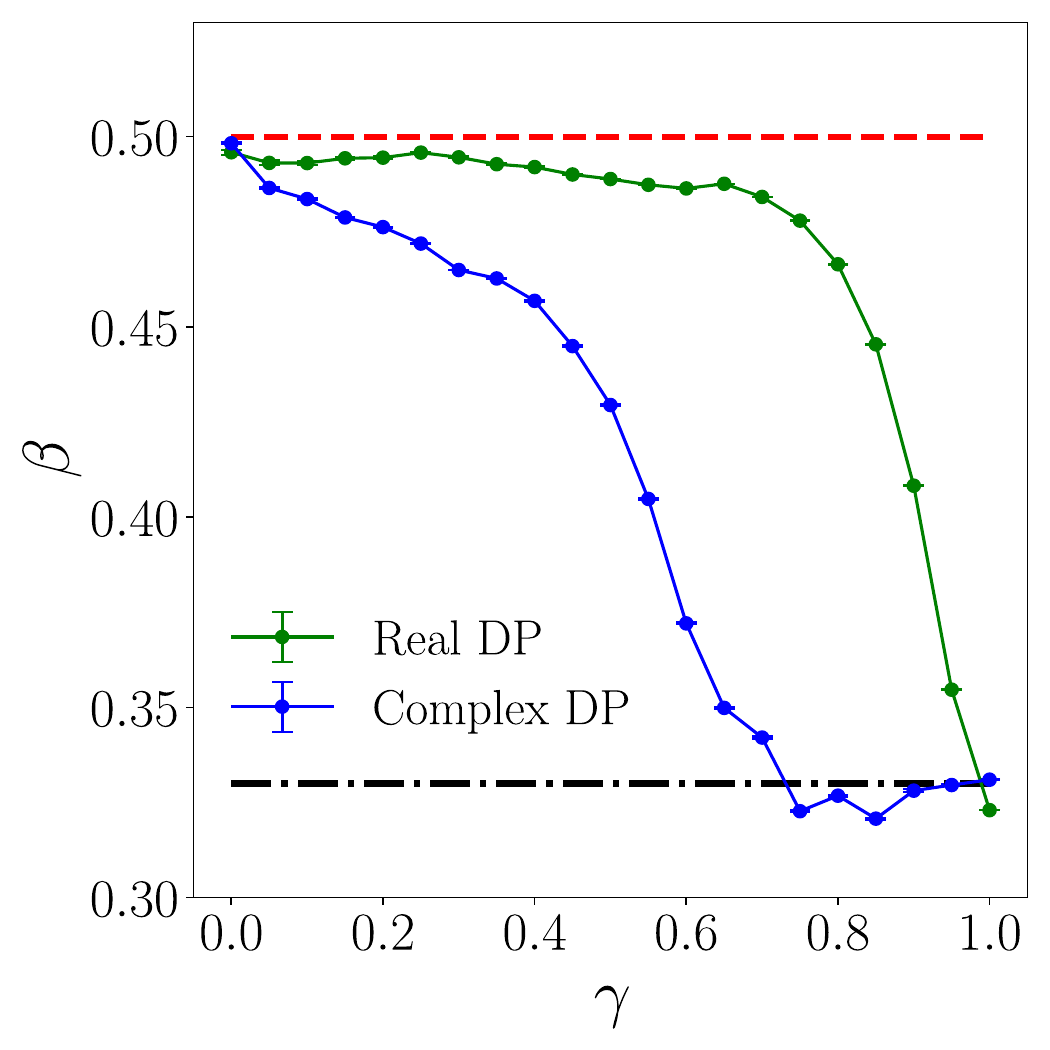}
\caption{Scaling of fluctuations in the free energy of real and complex directed polymer (DP), $\sigma[F(x,y)]\sim y^{\beta}$, considering different ratios between point and columnar disorder $\gamma$. The red dashed and black dotted dashed lines represent the algebraic behaviors $y^{1/2}$ and $y^{1/3}$, respectively. Numerical simulations were performed on complex DP using Eq.~\eqref{Eq:cDP_tranfer} and real DP with real weights for $x=0$, i.e. from the initial point $(0,0)$ to $(0,y)$, up to $y=4096$ with $W_V=W_D=\pi$ and $30000$ disorder realizations.}
\label{fig:dp_std_vs_t_p_c}
\end{figure}

{To validate the universality of the observed fluctuation scaling in the localized wavefunction, we consider the complex directed polymers (DP) in random media~\cite{Zhang_prl_complex_dp,Zhang_epl_complex_dp,Medina1989,GELFAND199167,BLUM1992588,Roux_1994}, examining how the competition between point and columnar disorder influences the scaling of free energy fluctuations.}

{While it is common to treat quenched disorder as uncorrelated point defects in real materials or toy models, studies have shown that columnar defects are more effective than point defects at pinning flux lines at high temperatures and fields in superconductors~\cite{Civale91prl,Hwa93prl}. This highlights the significance of correlated disorder in shaping the transport properties of the systems, leading to a variety of intriguing developments, e.g. mapping from the physics of flux lines to localization of interacting quantum particles in the Bose-glass phases~\cite{Nelson93prb,Giamarchi96prb}.
The localization (or delocalization) of real DP onto (or from) a columnar defect under competing point disorder has been examined in earlier studies as well~\cite{Tang1993prl,Balents1993epl}. When real DP are subjected to both point and columnar defects, it has also been observed that the positional fluctuations exceed those observed in systems with purely columnar or purely point disorder~\cite{Arsenin1994pre}. Furthermore, the behavior of real DP with spatially correlated disorder has been studied~\cite{Kardar1987,Medina1989pra,Chu2016pre}. When the disorder is described by the covariance:
\begin{equation}
{\rm Cov}[V(x, y), V(x', y')] \sim |x-x'|^{2\rho-1}\delta(y-y'),
\label{eq:corr_dis}
\end{equation}
the fluctuation scaling aligns well with predictions from a one-loop dynamical renormalization group calculation. This approach predicts the fluctuation growth exponent as a function of $\rho$~\cite{Forster77pra,Medina89pra}:
\begin{equation}
\beta(\rho) = \left\{ 
\begin{array}{ll} 
1/3, & 0 < \rho < 1/4, \\ 
(1+2\rho)/(5-2\rho), & 1/4 < \rho < 1. 
\end{array} 
\right.
\end{equation}
}

\subsection{DP with competing point and columnar disorder}
The DP described by Eq.~\eqref{eq:z_dp} is embedded in a disordered medium $V(x,y)$ (illustrated in Fig.~\ref{fig:fock_space_to_dp}) resulted from $ V_p(i, j) + V_c(i, j)$, where $V_p(i, j)$ and $V_c(i, j)$ represent the point and columnar disordered on-site energies defined on the Fock space (illustrated in Fig.~\ref{fig:fock_space}), respectively. The point disorder $V_p(i, j)$ is independent, identically distributed (i.i.d), and uniformly drawn from the interval $[-W_D/2, W_D/2]$. The columnar disorder $V_c(i, j)$, on the other hand, is a Cartesian sum of the same one-dimensional disorder: $V_c(i, j) = V_i + V_j$, where $V_i$ is i.i.d uniformly sampled from the interval $[-W_V/2, W_V/2]$. {The correlations in $V_c(i, j)$ can be characterized by the covariance:
\begin{eqnarray}
{\rm Cov}[V_c(i, j), V_c(k, l)] &=& \langle V_c(i, j)V_c(k, l)\rangle - \langle V_c(i, j)\rangle \langle V_c(k, l)\rangle \nonumber\\
&=& \frac{W^2_V}{12}(\delta_{i,k} + \delta_{i,l} + \delta_{j,k} + \delta_{j,l}).
\label{eq:cov}
\end{eqnarray}
In the Fock space of our TIP problem when $W_D=0$, the energies of all sites where one particle is at a fixed position $i^\star$ are always correlated, regardless of their distance. Specifically, ${\rm Cov}[V_c(i^\star, j), V_c(i^\star, l)] = W^2_V/12$, indicating a constant correlation independent of the separation between the two sites. This spatially invariant correlation gives rise to what is referred to a columnar disorder in our consideration.
}

{In the following, we consider the combined contribution of both types of disorder to the fluctuation scaling behavior,
\begin{equation}
\gamma V_p(i, j) + (1-\gamma) V_c(i, j) \rightarrow V(x,y),
\end{equation}
where $0 < \gamma < 1$ controls the relative strength of the point disorder $V_p(i, j)$ and the columnar disorder $V_c(i, j)$ in the resulted $V(x,y)$ for the DP problem. Note that $V(x, y)$ is derived from $\gamma V_p(i, j) + (1-\gamma) V_c(i, j)$ via a $\pi/4$ rotation in the coordinate system, as illustrated in Fig.~\ref{fig:fock_space} and Fig.~\ref{fig:fock_space_to_dp}.
Consequently, $V(x, y)$ directly inherits the correlations of $V_c(i, j)$ when $\gamma\neq 1$. For each disorder realization, we compute the contributions of directed paths to $Z(x, y)$ using a transfer matrix method~\cite{Kardar_prl1987,mezard1990glassy,Kardar1992,Zhang1995dp}. $Z(x, y+1)$ depends only on $Z(x-1, y)$, $Z(x+1, y)$, and the onsite disorder $V(x, y+1)$, following:
\begin{equation}
Z(x, y+1) = e^{-i V(x, y+1)} \left(Z(x-1, y) + Z(x+1, y)\right).
\label{Eq:cDP_tranfer}
\end{equation}
In this complex DP model, we identify $|Z(x, y)|^2$ as the partition function, and the free energy is defined as $F(x, y) = -\ln{|Z(x, y)|^2}$. The standard deviation of the free energy $\sigma[F(x, y)]$ scales as a function of $y$ according to 
\begin{equation}
\sigma[F(x, y)] \sim y^\beta,
\end{equation} 
with the fluctuation growth exponent $\beta$. In the case when $\gamma=1$ (i.e., only point disorder), the complex DP also belongs to the KPZ universality class with the exponent $1/3$~\cite{Zhang_prl_complex_dp,Zhang_epl_complex_dp,Medina1989,GELFAND199167,BLUM1992588,Roux_1994,PhysRevLett.111.026801}.} 

\subsection{Numerical results on the free energy fluctuation}

We now examine numerically how the competition between point and columnar disorder influences the scaling of free energy fluctuations in the complex DP model. Our results for the scaling of free energy fluctuations are presented in Fig.~\ref{fig:dp_std_vs_t_p_c}, where we explore different ratios of point to columnar disorder. We clearly observe a crossover in the fluctuation growth exponent from $1/2$ to $1/3$ as the columnar disorder is reduced and the point disorder is correspondingly increased.
These results closely align with our previous findings on the scaling of fluctuations in $\ln|\psi_{\rm loc}(r)|^2$ as a function of the graph distance $r$ in the Fock space, obtained from varying the relative strengths of random long-range interaction and random potential. This agreement confirms that complex DP with columnar disorder maps to the localized wavefunction in the Fock space of TIP with weak or short-range interacting cases, while point disorder corresponds to the strong random long-range interactions.

{Specifically, when only $V_c(i, j)$ contributes to $V(x, y)$ (i.e., $\gamma = 0$), the polymer can become pinned to a particular column corresponding to a fixed $i^\star$ or $j^\star$ for an extended length~\cite{Kardar1987,Medina1989pra}. Along the pinned column, the energy $V(i, j)$ is determined by $V_{i^\star} + V_j$ or $V_i + V_{j^\star}$, respectively, leading to fluctuations scaling similar to a random walk. This long-range correlation profoundly influences the fluctuation scaling behavior, distinguishing the columnar disorder from point disorder. In contrast, the pinning of the polymer in the presence of point disorder arises from a global optimization process, reflecting the fundamentally different nature of these two types of disorder. Besides, we remark that our system at $\gamma=0$ resembles the scenario with $\rho = 1/2$ in Eq.~\eqref{eq:corr_dis} for persistent spatial correlations. The partition function is then the addition of random variables across different $y$ values. This leads to a fluctuation scaling in $y$ with an exponent of $1/2$. For our TIP system with only random potential, the fluctuation scaling aligns with our earlier claim that it exhibits behavior analogous to one-dimensional Anderson localization~\cite{Mirlin_pr2000}.}

This analysis provides insights into the structure of disordered on-site energies in the Fock space of the TIP through the lens of fluctuation scaling. We demonstrate that the effective tight-binding model defined in the Fock space is fundamentally distinct from standard two-dimensional Anderson localization, owing to the strong correlations introduced by the random potential in the absence of random long-range interactions. The mapping between physical disorder types and the disorder structure in the complex DP model further supports the universality of the observed fluctuation scaling. Notably, our numerical simulations for real DPs reveal a similar crossover (or transition), in which the fluctuation growth exponent of $1/3$ emerges only under purely point disorder ($\gamma=1$), as shown in Fig.~\ref{fig:dp_std_vs_t_p_c}. Noting that correlated disorder (columnar disorder) and uncorrelated disorder (point disorder) lead to fundamentally different fluctuation scalings in the localized wavefunction, we emphasize the importance of recognizing the structure of the disorder for developing accurate theoretical frameworks to describe disordered quantum interacting systems.

\section{Conclusion}
\label{conclusion}

In this work, we explored the universal fluctuation of localized wavefunction in the Fock space of TIP in strongly disordered one-dimensional systems. By using the forward scattering approximation, we mapped the localized wavefunction to the partition function of a DP problem in random media, and we examined how the interplay between random potentials and random long-range interactions affects fluctuation scaling. Our analysis revealed that random potentials alone induces strong columnar correlations in the on-site energies in the Fock space, resulting in fluctuation scaling characterized by the exponent $\beta \approx 1/2$. Introducing random long-range interactions disrupts these correlations, giving rise to point disorder and transitioning the fluctuation scaling to the KPZ universality class with $\beta \approx1/3$.

To validate the universality of these results, we studied a complex DP model with competing point and columnar disorder. The agreement between the fluctuation scaling in the localized wavefunction of TIP in the Fock space and the complex DP model confirmed the correspondence between random potentials and columnar disorder, as well as between random long-range interactions and point disorder. This mapping sheds light on the nontrivial role of disordered interactions and disorder correlations in shaping the universality class of fluctuation scaling in localized quantum systems.

{Through the lens of universal fluctuations, our findings show that disordered quantum interacting systems without random long-range interactions in the Fock space differ fundamentally from standard Anderson localization in higher dimensions~\cite{Tikhonov16,Lemarie17prl}, owing to the correlations of the on-site energies. By contrasting correlated (columnar) and uncorrelated (point) disorder, we underscore how the structure of disorder crucially shapes the universal fluctuations of the localized wavefunction and remark the pinning of the directed polymer by columnar disorder. When more particles are considered ($N > 2$) up to half filling, and in the absence of random long-range interactions, the on-site energies in the Fock space become even more strongly correlated~\cite{Roy20prl,Welsh_2018,mirlin2024prb}. This is because the system size $L$ remains fixed while the number of Fock-space sites increases as $N_F = \binom{L}{N}$. Even the inclusion of random two-body long-range interactions cannot remove these correlations completely, since the number of pairwise interactions, ${L(L-1)}/{2}$ is still smaller than the total number of sites in the Fock space. Consequently, an important challenge remains in understanding how such correlated on-site energies influence the fluctuations of the localized wavefunction in strongly disordered regimes, thereby opening new directions for both theoretical and experimental studies. These insights could offer a complementary perspective on the multifractal properties of many-body wavefunctions and the MBL transition in the Fock space~\cite{Laflorencie19,Tarzia2020prb,Laflorencie21,Roy_2024}. In particular, it could also shed new light on the statistics of rare long-range resonances in many-body eigenstates~\cite{Biroliprb2024} which is important to understand the critical properties of the MBL transition and its destabilization through quantum avalanches.}

\section{Acknowledgment}

S.M. thanks Wen Wei Ho, Sthitadhi Roy and Anushya Chandran for helpful discussions.  This work is supported by the Singapore National Research
Foundation via Project No.~NRF2021-QEP2-02-P09, as well as by the National Research Foundation, Singapore, and A*STAR under its CQT Bridging Grant. This study was supported by research funding Grants No. ANR-19-CE30-0013, and by the Singapore Ministry of Education Academic Research Funds Tier II (WBS No. A-8001527-02-00 and A-8002396-00-00). We thank Calcul en Midi-Pyr\'en\'ees (CALMIP), the National Supercomputing Centre (NSCC) of Singapore and MPIPKS for computational resources.

\bibliography{reference.bib}

\end{document}